\documentstyle[aps]{revtex}

\newcommand{\ds}{\displaystyle}

\newcommand{\non}{\nonumber}

\newcommand{\beq}{\begin{eqnarray}}
\newcommand{\eeq}{\end{eqnarray}}

\newcommand{\be}{\begin{eqnarray}}
\newcommand{\en}{\end{eqnarray}}
\newcommand{\spin}{{\vec S}}

\newcommand{\JPSJ}{J. Phys. Soc. Jpn.}

\newcommand{\PRL}{ Phys. Rev. Lett.}
\newcommand{\PRB}{ Phys. Rev. {\bf  B}}

\newcommand{\PROG}{Prog. Theor. Phys.}
\begin{document} 
\draft 
\title{\vspace*{2.5cm} Effects of dimerization  and interchain one-particle hopping  in
a weakly coupled dimerized chain system at quarter filling}
\author{J. Kishine and K. Yonemitsu} 
\address{ Institute for Molecular Science, Okazaki 444-8585, Japan \\ 
{\rm \footnotesize (submitted to Synthetic Metals, proceedings of International
 Conference on Synthetic Metals, 1998, Montpellier, France)}\\
{\rm \vspace{12pt}\parbox{17.6cm}{ 
\line(1,0){176} \\ 
{\bf Abstract}\\ \hspace*{1em}
Effects of the intrachain dimerization and     
the interchain one-particle hopping, $t_{b}$, in  a quasi-one-dimensional dimerized
chain system at quarter filling have been studied, based on the perturbative renormalization group (PRG) approach.
Based on the results, we discuss   difference in the low-energy properties between    TMTTF and TMTSF compounds.
\\ \line(1,0){176} 
}}}
\maketitle
\baselineskip16pt
\section{Introduction}

Quasi-one-dimensional (Q1D) organic conductors (TMTTF)$_2X$ and (TMTSF)$_2X$ ( $X$=Br, PF$_6$,...) have  in common
the 2:1 stoichiometry, which makes the band quarter-filled, and the   dimerized
one-particle hopping integrals, $t_{a1}$ and $t_{a2}$,
 along the conducting stack\cite{JeromeSchulz}. 
The ratios of the one-particle hopping integrals along the high and  intermediate  directions, $t_a$ and $t_b$,
 are approximately $t_{b}/t_{a}\sim 0.04$ and 0.1 for TMTTF and TMTSF compounds, respectively.

In spite of the similarity in  electronic structure,     
 their low temperature transport and magnetic properties are essentially different.
At ambient pressure, (TMTTF)$_2$Br is semiconducting with a shallow minimum in resistivity 
and undergoes a phase transition    to a commensurate spin-density wave (CSDW) phase\cite{Takahashi86}.
In contrast, (TMTSF)$_2$PF$_6$ shows  metal-like behavior down to a phase transition   to
 an incommensurate SDW (ICSDW)\cite{Nakamura95}.
Optical reflectivity spectra  above the phase transition temperature 
indicate  the   one-particle propagation is {\it confined} in a single chain  at any temperature for (TMTTF)$_2$Br, 
but  it is
{\it deconfined }   at low temperatures for  (TMTSF)$_2$PF$_6$\cite{Gruner}.

As was stressed by Emery {\it et al.}\cite{Emery82}, important difference   between TMTTF and TMTSF compounds is  
the degree of   dimerization,
 $\Delta=(t_{a1}-t_{a2})/(t_{a1}+t_{a2})$,    evaluated as 0.2 and 0.05, 
respectively\cite{Mila95}.
In terms of $g$-ology\cite{Solyom}, the intrachain backward, forward and $2k_F$-umklapp scattering strengths  in a 
dimerized chain at quarter filling are given by
$\pi v_F g_1=U/2-V$,
$\pi v_F g_2=U/2+V$, 
$\pi v_F g_3=\left({U/ 2}-V\right)\ds{2 \Delta/( 1+\Delta^2)}$\cite{PencMila94},
where $g_i$ are  dimensionless scatering strengths with $v_F$ being the Fermi velocity and $U$, $V$ denoting the on-site and 
the nearest neighbor Coulomb
repulsions, respectively. 
Therefore  stronger  $\Delta$ causes   stronger $g_3$.
Recently, based on the perturbative renormalization group (PRG) approach\cite{Bourbonnais91},   we discussed the
 effects of the umklapp scattering 
in   a weakly-coupled {\it half-filled} chain system\cite{JK}. 
 We  here extend the work to the case of
 a weakly coupled   dimerized  chain system at  quarter filling  where a finite dimerization causes a finite umklapp 
scattering strength.

\section{PRG Formulation}

We consider a 2D array of an infinite number of chains weakly coupled via   interchain one-particle hopping $t_b$. 
As in Ref.\cite{JK}, we treat   renormalization flows of  $g_i$ by solving the 2-loop PRG equations\cite{Kimura75}.
When the initial values of $g_i$ satisfy the condition 
$
g_{1}-2g_{2}< \mid g_{3}  \mid , 
$
 the umklapp process becomes relevant
and the 2-loop RG equations give the non-trivial fixed point, $g_1^\ast=0$ 
and $\mid g_{3}^{\ast}\mid=2g^{\ast}_{2}-g_{1}^{\ast}=2$.
From now on, we   consider only  this parameter region. 
In the absence of the interchain coupling,  the fixed point corresponds to   
 the Mott insulator phase with short-range antiferromagnetic (AF) 
 correlation.

The 2-loop PRG  equation for $t_b$ is given by\cite{Kimura75,Bourbonnais93}
%%%%%%%%%%%%%%%%%%%%%%%%%%%%%%%%%%%%%%%%%%%%%%%%%%%%%%%%%%%%%%%%%%%%%%%%
\begin{eqnarray}
{d  \ln t_b/ dl}=1-\left({g_1}^2+{g_2}^2-{g_1}{g_2}+{g_3}^2/2\right)/4, \label{eqn:t}
\end{eqnarray}
%%%%%%%%%%%%%%%%%%%%%%%%%%%%%%%%%%%%%%%%%%%%%%%%%%%%%%%%%%%%%%%%%%%%%%%%
where   the scaling parameter, $l$, is related to the absolute temperature, $T$, as   $l=\ln[E_0/T]$ with
the high-energy bandwidth cutoff, $E_0$, which is of the order of the intrachain hopping integral, $t_a$.
In the course of the renormalization, $t_b$ attains the order of  the initial bandwidth, $E_{0}$, 
at some crossover value of the scaling parameter, $l_{\rm cross}=\ln [E_{0}/ T_{\rm cross}]$, 
qualitatively defined by
$
t_{b}(l_{\rm cross})= E_{0}.\label{eqn:defopc}
$

In the paramter region considered here, the most dominant interchain two-particle 
process {\it dynamically generated} in the course of
the scaling\cite{Bourbonnais91} is the interchain $2k_F$ spin-spin interaction.
The corresponding interaction Hamiltonian is written as
\begin{equation}
\!\!\!\!\!\!\!\!\!\!\!\!\!\!\!
{\cal H}_{\perp}^{\rm int}\!\!=\!\!{\pi v_{F}\over 4 }\sum_{\vec q}\left[
J(q_b)
\spin^{\ast}_{\vec q}\cdot\spin_{\vec q}
+K(q_b)(\spin^{\ast}_{\vec q}\cdot\spin^{\ast}_{\vec q}+\spin_{\vec q}\cdot\spin_{\vec q})\right],  
\end{equation}
where $\spin_{\vec q}$ denotes the $2k_F$ spin density field with the momentum 
$\vec q=(2k_F, q_b)$ ( $q_b$ denotes  the momentum  perpendicular to the chain).

The PRG  equations for the interchain spin-spin interactions are written as
%%%%%%%%%%%%%%%%%%%%%%%%%%%%%%%%%%%%%%%%%%%%%%%%%%%%%%%%%%%%%%%%%%%%%%%%
\begin{eqnarray}
d J(q_b)/ dl&=&{1\over 2}{ \tilde t_{b}}^2\left[{g_2}^2+4{g_3}^2\right]\cos q_b  
+{1\over 2}\left[g_2 J(q_b)+4g_3 K(q_b)\right]
-{1\over 4}\left[J(q_b)^2+4K(q_b)^2\right],\label{eqn:J}\non \\
{d K(q_b)/dl}&=&2{ \tilde t_{b}}^2
{g_2}{g_3}\cos q_b 
+2\left[g_2 K(q_b)+g_3 J(q_b)\right]-J(q_b) K(q_b)\label{eqn:K},
\end{eqnarray}
%%%%%%%%%%%%%%%%%%%%%%%%%%%%%%%%%%%%%%%%%%%%%%%%%%%%%%%%%%%%%%%%%%%%%%%%
where $\tilde t_{b}\equiv t_{b}/ E_0$.
Although     $J(q_b)=K(q_b)=0$ at the initial step, the third term  causes divergence of them at a 
critical scaling parameter $l_{\rm SDW}=\ln [E_{0}/ T_{\rm SDW}]$ defined by
$
J(q_b)=K(q_b)=-\infty\label{eqn:defoflc2}.
$
The value of $l_{\rm SDW}$ becomes minimum for $q_b=\pi$ corresponding to   
a 2D commensurate SDW phase(i.e., 2D AF  phase). 
From now on,  we fix $q_b=\pi$  and replace $T_{\rm SDW}$ with  the Neel temperature, $T_{N}$.

\section{Phase Diagram}

To see which of $T_{\rm cross}$ and $T_{N}$ is larger, we solve the coupled scaling equations (\ref{eqn:t}), 
  (\ref{eqn:J}) and (\ref{eqn:K}).
We treat the RG flows of   $g_i$ through the 2-loop PRG equations.
As  to the   initial conditions  for the intrachain Coulomb repulsions, we use  $U=4V=1.6 \pi v_F$.

In Fig.~1, we show a phase diagram  spanned by $t_{b0}$ (the initial value of
the interchain one-particle hopping integral) and $T$
for $\Delta=0.2$.
 There exists a critical value of $t_{b}$: $t_{b}^{\ast}\sim 0.23 E_0$.
For $t_{b} < t_{b}^{\ast}$, the interchain  one-particle propagation is strongly suppressed  and 
the 2D AF phase is stabilized at $T_{N}$. 
For   $t_{b} > t_{b}^{\ast}$, the interchain  one-particle propagation develops and
the system undergoes a crossover to  the Fermi liquid (FL)  phase. 
 In the FL phase,  the phase transition to the SDW phase  due to the Fermi surface nesting is possible, where
the SDW vector  is 
determined by the optimal nesting condition which generally leads to the ICSDW transition.
In this case,  the increasing $t_b$ decreases the degree of   nesting and consequently decreases the SDW 
transition temperature\cite{Yamaji83} and finally a superconducting
transition is caused by the spin fluctuation mechanism\cite{SLH87}.

In Fig.~2, we show how  $t_{b}^{\ast}$  
depends on  $\Delta$.  
We see that a finite $\Delta$ causes a finite $ t_{b}^{\ast}$.
As $\Delta$ becomes weaker,  
  the  AF phase in Fig.~1  shrinks. This situation comes from the fact that   the  umklapp scattering
 becomes less important and the suppression 
of the interchain one-particle hopping becomes weaker with the decreasing $\Delta$, and consequently 
the interchain one-particle propagation
can acquire   coherence even for small $t_b$.
Provided that  TMTTF and TMTSF compounds have the same strength of the intrachain interaction, $U=4V=1.6 \pi v_F$,
$t_b$ and $\Delta$  of TMTTF and TMTSF compounds  are
  located at the points indicated in Fig.~2.
We see the point of TMTTF lies in the AF region, while the points of TMTSF lies near the boundary between the AF and the FL phase.

\section{Summary}

Based on the PRG approach, we have studied the effects of the dimerization, $\Delta$, and the interchain one-particle
 hopping integral, $t_b$, in
the weakly coupled dimerized chain system at quarter filling. A finite dimerization causes a finite strength of 
the intrachain umklapp scattering and
consequently the interchain one-particle propagation is strognly suppressed. Then the low-energy asymptotics of 
the system is determined through $\Delta$ and
$t_b$. The present results qualitatively explain the difference in the low-energy properties between TMTTF and 
TMTSF compounds.
 J.K was   supported by a Grant-in-Aid for Encouragement of Young Scientists   from the Ministry of Education,
 Science, Sports and Culture, Japan.

\begin{figure}
\caption{
Phase diagram.
{\bf AF} and {\bf FL} are the abbreviations for the antiferromagnetic phase and the Fermi liquid phase, respectively. 
}
\end{figure}

\begin{figure}
\caption{
Dependence of $t_{b}^\ast$ on $\Delta$. 
$t_b$ and $\Delta$  of TMTTF and TMTSF compounds  are indicated.
}
\end{figure}


\noindent
\begin{references}
\vspace{-46pt}
\bibitem{JeromeSchulz} D. Jerome and H. J. Schulz, Adv. Phys. {\bf 31}  (1982)  229.
\bibitem{Takahashi86} T. Takahashi  {\it et al.},  {\JPSJ} {\bf 55} (1986) 1364.
\bibitem{Nakamura95} T. Nakamura {\it et al.}, Synth. Met. {\bf 70} (1995) 1293.
\bibitem{Gruner} V. Vescoli  {\it et el.}, submitted to Science.
\bibitem{Emery82}V.J.Emery, R. Bruinsma and S. Barisic,  {\PRL} {\bf 48} (1982) 1039.
\bibitem{Mila95} F. Mila, {\PRB} {\bf 52} (1995) 4788.
\bibitem{Solyom}J. S\'olyom,  Adv. Phys. {\bf 28} (1979) 201.
\bibitem{PencMila94} K. Penc and F. Mila, {\PRB} {\bf 50} (1994)11429.
\bibitem{Bourbonnais91}C. Bourbonnais and L. G. Caron,  Int. J. Mod.  Phys. {\bf B5}  (1991)  1033.
\bibitem{JK} J. Kishine and K. Yonemitsu, to appear in {\JPSJ} {\bf 67} (1998) No. 8 (cond-mat/9802186).
\bibitem{Kimura75}M. Kimura,  {\PROG} {\bf 53} (1975) 955.
\bibitem{Bourbonnais93}C. Bourbonnais,  J.Phys. I France {\bf 3} (1993) 143.
\bibitem{Yamaji83} K. Yamaji,  {\JPSJ} {\bf 52} (1983) 1361.
\bibitem{SLH87} D. J. Scalapino, E. Loh, Jr. and J. E.  Hirsh,  {\PRB} {\bf 34} (1986)8190.
\end{references}
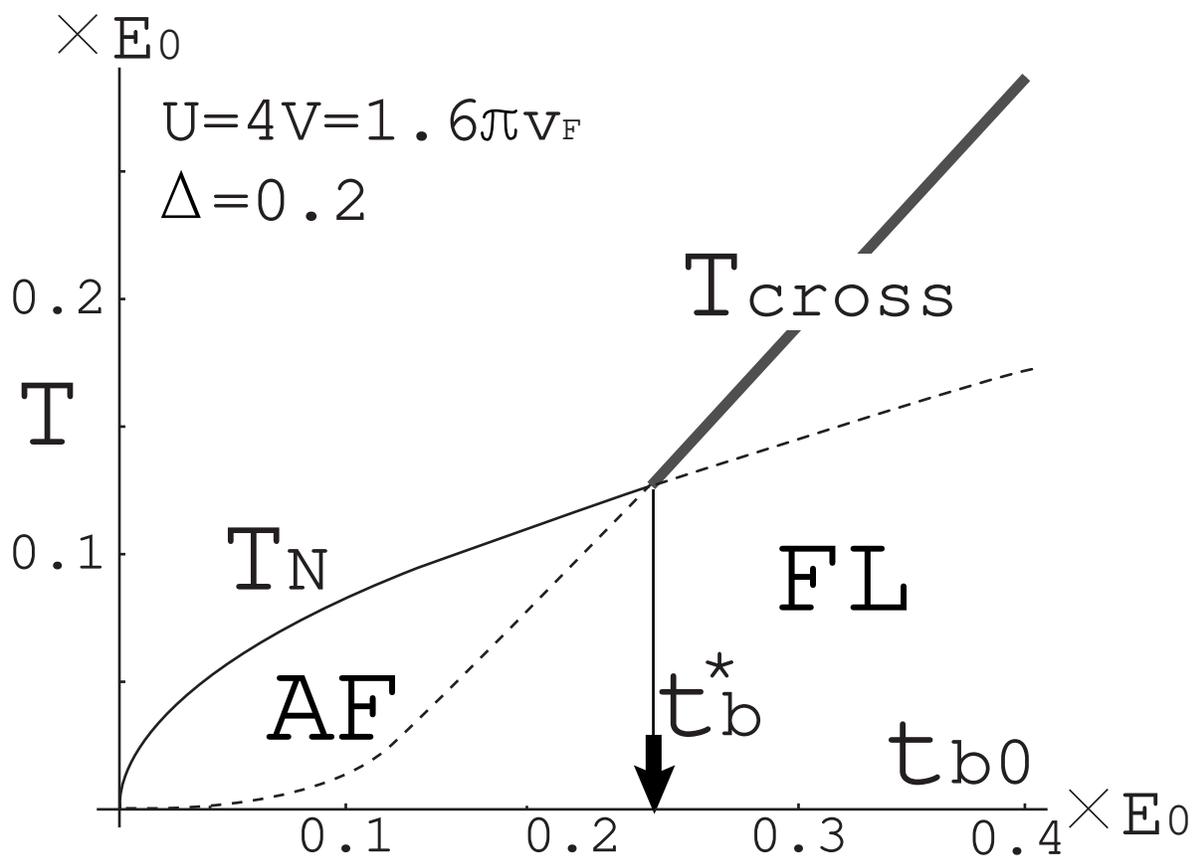
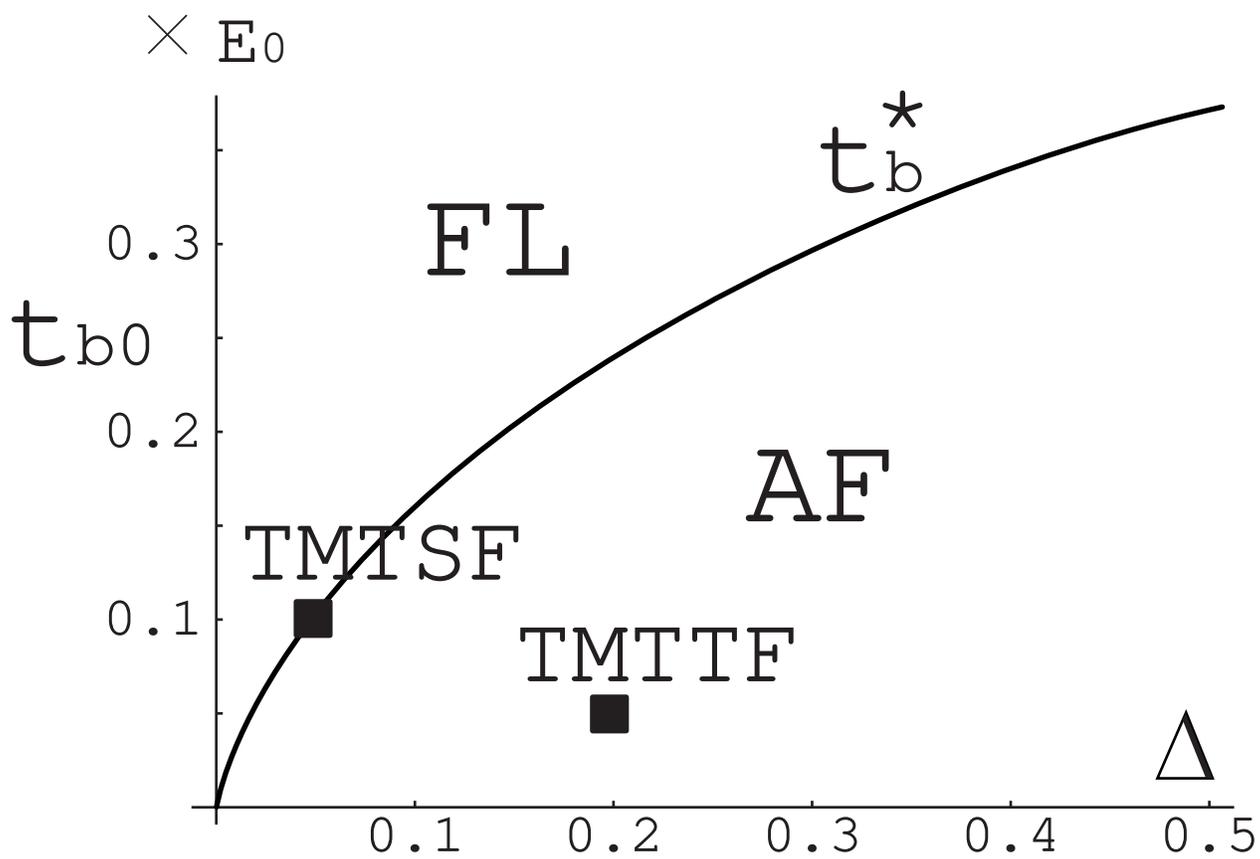
\end{document}